\documentclass[]{spie}  

\usepackage{amsmath,amsfonts,amssymb}
\usepackage{graphicx}
\usepackage[colorlinks=true, allcolors=blue]{hyperref}

\title{The high-contrast performance of the Keck Planet Imager and Characterizer}

\author[a,k]{Jason J. Wang}
\author[b,c]{Dimitri Mawet}
\author[b]{Jerry W. Xuan}
\author[a]{Chih-Chun Hsu}
\author[d]{Jean-Baptiste Ruffio}
\author[b]{Katelyn Horstman}
\author[b]{Yinzi Xin}
\author[e]{Jacques-Robert Delorme}
\author[b]{Nemanja Jovanovic}
\author[b,$\star$]{Yapeng Zhang}
\author[f]{Luke Finnerty}
\author[b]{Ashley Baker}
\author[c]{Randall Bartos}
\author[j]{Geoffrey A. Blake}
\author[f,b]{Benjamin Calvin}
\author[e]{Sylvain Cetre}
\author[e]{Gregory W. Doppmann}
\author[b]{Daniel Echeverri}
\author[f]{Michael P. Fitzgerald}
\author[g,b]{Joshua Liberman}
\author[f]{Ronald Lopez}
\author[h]{Evan Morris}
\author[b]{Jacklyn Pezzato-Rovner}
\author[d]{Ben Sappey}
\author[b]{Tobias Schofield}
\author[h]{Andrew Skemer}
\author[c]{J. Kent Wallace}
\author[i]{Ji Wang}
\affil[a]{Center for Interdisciplinary Exploration and Research in Astrophysics (CIERA), Northwestern University,
1800 Sherman Ave, Evanston, IL, 60201, USA}
\affil[b]{Department of Astronomy, California Institute of Technology, Pasadena, CA 91125, USA}
\affil[c]{Jet Propulsion Laboratory, California Institute of Technology, 4800 Oak Grove Dr.,Pasadena, CA 91109, USA}
\affil[d]{Department of Astronomy and Astrophysics, University of California, San Diego, La Jolla, CA 92093}
\affil[e]{W. M. Keck Observatory, 65-1120 Mamalahoa Hwy, Kamuela, HI, USA}
\affil[f]{Department of Physics \& Astronomy, 430 Portola Plaza, University of California, Los Angeles, CA 90095, USA}
\affil[g]{James C. Wyant College of Optical Sciences, University of Arizona, Meinel Building 1630 E. University Blvd., Tucson, AZ. 85721, USA}
\affil[h]{Department of Astronomy \& Astrophysics, University of California, Santa Cruz, CA 95064, USA}
\affil[i]{Department of Astronomy, The Ohio State University, 100 W 18th Ave, Columbus, OH 43210 USA}
\affil[j]{Division of Geological and Planetary Sciences, California Institute of Technology, Pasadena CA 91125, USA}
\affil[k]{Department of Physics and Astronomy, Northwestern University, 2145 Sheridan Rd, Evanston, IL 60208, USA}

\authorinfo{$^\star$51 Pegasi b Fellow. \\ Further author information: (Send correspondence to J.J.W.: E-mail: jason.wang@northwestern.edu}

\begin{document} 
\maketitle

\begin{abstract}
The Keck Planet Imager and Characterizer (KPIC), a series of upgrades to the Keck II Adaptive Optics System and Instrument Suite, aims to demonstrate high-resolution spectroscopy of faint exoplanets that are spatially resolved from their host stars. In this paper, we measure KPIC’s sensitivity to companions as a function of separation (i.e., the contrast curve) using on-sky data collected over four years of operation. We show that KPIC is able to reach contrasts of $1.3 \times 10^{-4}$ at 90 mas and $9.2 \times 10^{-6}$ at 420 mas separation from the star, and that KPIC can reach planet-level sensitivities at angular separations within the inner working angle of coronagraphic instruments such as GPI and SPHERE. KPIC is also able to achieve more extreme contrasts than other medium-/high-resolution spectrographs that are not as optimized for high-contrast performance. We decompose the KPIC performance budget into individual noise terms and discuss limiting factors. The fringing that results from combining a high-contrast imaging system with a high-resolution spectrograph is identified as an important source of systematic noise. After mitigation and correction, KPIC is able to reach within a factor of 2 of the photon noise limit at separations $<$ 200 mas. At large separations, KPIC is limited by the background noise performance of NIRSPEC.
\end{abstract}

\keywords{Exoplanets, instrumentation, high-resolution spectroscopy, high-contrast imaging, Keck telescope}

\section{INTRODUCTION}
\label{sec:intro}  

The direct imaging of exoplanets that are spatially resolved from their host stars is difficult because the diffracted light of the star is often times orders of magnitude brighter than the exoplanets themselves. Due to changing atmospheric turbulence and drifts in optical alignment, high-contrast instruments cannot maintain a perfectly static diffraction pattern of the star. When using images of the diffraction pattern taken at other times to subtract off the stellar light, the removal is therefore imperfect, and has been shown to limit sensitivity at close angular separations from the star \cite{Bailey2016}. 

The combination of high-contrast imaging with high-resolution spectroscopy (HCI+HRS) offers a way to circumvent the imperfect subtraction of diffracted starlight. The approach works by feeding the light of the companion and light diffracted from the star coming from the high-contrast imaging system into a high-resolution spectrograph ($R > 10,000$). At these spectral resolutions, we can resolve individual spectral lines and use the fact that the light of the planet has molecular absorption or emission lines that not present in the stellar spectrum and that are Doppler-shifted relative to the star due to its orbital motion. This has been shown to work well for both detecting spatially unresolved planets in combined light \cite{Birkby2013,Line2021,Finnerty2023} as well as for providing the extra gain in sensitivity needed to detect spatially resolved planets buried underneath the glare of the star \cite{Snellen2014,Hoeijmakers2018,Agrawal2023}. The latter has generally been done with spectrographs that were not originally conceived nor optimized for high-contrast performance. However, such optimized instruments are critical in order to reach down to the sensitivities of terrestrial planets in the habitable zones of other stars when coupled with the next generation of extremely large telescopes \cite{Kasper2021,Mawet2022,Kautz2023}. The sensitivities needed require HCI+HRS instruments to reach down to photon noise limits and not to be limited by systematics in the removal of diffracted starlight \cite{WangJi2017}.

The Keck Planet Imager and Characterizer (KPIC) is one of the first such instruments that is designed from the beginning to be optimized for HCI+HRS \cite{Mawet2017,Delorme2021,Echeverri2022}. It feeds the light from the Keck II adaptive optics (AO) system into the existing NIRSPEC spectrograph \cite{McLean1998,Martin2018,Lopez2020} through a set of four single-mode fibers that sample different spatial locations in the AO field-of-view. The fiber injection unit (FIU) is used to steer the input so that the light of companion hidden underneath the diffracted starlight can be coupled into the spectrograph \cite{Delorme2021,Wang2021}. The single-mode fiber not only provides passive starlight rejection, but active wave front control strategies can be used with them to further mitigate the amount of starlight that is transmitted to the spectrograph \cite{Mawet2017,Xin2023}. Since 2020, KPIC has been able to detect and characterize the atmospheres, spins, and orbits of known directly imaged exoplanets \cite{Wang2021}. In this work, we use the first four years of KPIC performance to assess its sensitivity limits and identify the sources of noise that limit the HCI+HRS technique on KPIC. 

\section{OBSERVATIONS AND DATA REDUCTION}

\subsection{Observations}
The angular extent of each single mode fiber in KPIC is matched to the size of Keck's diffraction-limited point spread function (PSF). Each KPIC observation consists, essentially, of probing a single spatial location relative to the star. We note that KPIC does have 4 fibers, so we actually sample 4 distinct spatial positions. However, most of these ``sky" positions are far from the star, so we choose to ignore the other fibers for this analysis. Instead, we focus on the fiber that is pointed at the position of a point source of interest. In order to obtain sufficient sampling for various angular separations from the star, we utilize the multiple years of observations obtained by KPIC as part of the spectroscopic survey it is undertaking. We select the datasets where weather conditions were good and the host star was bright, in order to assess the limits of the high-contrast performance. We list all the observations considered for this sensitivity analysis in Table \ref{tab:obs}. While the exposure times varied, they are long enough to assess the instrument performance and only range between 100 and 300 minutes total. For observations starting in 2021, we have used a nodding technique, where the light from the companion is injected into two fibers in an alternating fashion. This helps with more accurate subtraction of the instrumental thermal background. 

For all observations, spectra of the host star are taken at least once an hour for modeling the speckle light that leaks into the planet spectra. Observations of a standard star are taken right before or after each science sequence to measure the product of telluric transmission and instrument response. Another standard star (typically an M-giant) is observed each night to compute the wavelength solution. Lastly, one of the bright telluric standard stars is observed in all four science fibers in order to measure the position and width of each dispersed fiber trace across the detector. For the two observations of HD 206893 c, we ran speckle nulling routines to minimize the amount of stellar light coupling into the fiber \cite{Xin2023}.

\begin{table}[ht]
\caption{KPIC Observations Used in this work. All stellar Kmags are from \citenum{Cutri2003} and separations are from \citenum{witp}. } 
\label{tab:obs}
\begin{center}       
\begin{tabular}{|l|l|c|c|c|c|c|c|} 
\hline
\rule[2ex]{0pt}{3.5ex}  Date & Target & \shortstack[c]{Star\\Kmag} & \shortstack[c]{Planet\\Kmag} & \shortstack[c]{Separation\\(mas)} & \shortstack[c]{Int. Time\\(minutes)} & Detection? & Ref. \\
\hline
\rule[-1ex]{0pt}{3.5ex}  2020-07-01 & HR 8799 c & 5.240 & 15.9 & 951 & 190 & Yes & \citenum{Zurlo2016,Wang2021}   \\
\hline
\rule[-1ex]{0pt}{3.5ex}  2020-07-02 & HR 8799 d & 5.240 & 15.9 & 689 & 230 & Yes & \citenum{Zurlo2016,Wang2021}  \\
\hline
\rule[-1ex]{0pt}{3.5ex}  2020-07-03 & HR 8799 e & 5.240 & 15.9 & 398 & 110 & Yes & \citenum{Zurlo2016,Wang2021}  \\
\hline
\rule[-1ex]{0pt}{3.5ex} 2020-09-28 & HR 8799 & 5.240 & - & 200 & 120 & No & -  \\
\hline
\rule[-1ex]{0pt}{3.5ex} 2021-10-21 & HR 8799 b & 5.240 & 17.0 & 1721 & 280 & Yes & \citenum{Zurlo2016,Wang2021}  \\
\hline 
\rule[-1ex]{0pt}{3.5ex} 2022-07-20 & HD 206893 B & 5.593 & 14.9 & 197 & 140 & Yes & \citenum{Delorme2017}  \\
\hline 
\rule[-1ex]{0pt}{3.5ex} 2022-11-13 & HD 206893 c & 5.593 & 16 & 89 & 238 & No & \citenum{Hinkley2023}  \\
\hline 
\rule[1ex]{0pt}{3.5ex} \shortstack[c]{2023-06-16\\2023-06-21} & HIP 99770 b & 4.422 & 15.7 & 420 & 290 & Yes & \citenum{Currie2023} \\
\hline 
\rule[-1ex]{0pt}{3.5ex} 2023-06-29 & GJ 758 B & 4.493 & 18.7 & 1359 & 240 & No & \citenum{Vigan2016} \\
\hline 
\rule[-1ex]{0pt}{3.5ex} 2024-05-23 & HD 206893 c & 5.593 & 16 & 50 & 109 & No & \citenum{Hinkley2023} \\
\hline 
\end{tabular}
\end{center}
\end{table}

\subsection{Data Reduction}\label{sec:drp}
The data reduction for each target follows the procedure outlined in \citenum{Wang2021} and utilizes the KPIC Data Reduction Pipeline\footnote{https://github.com/kpicteam/kpic\_pipeline}. We briefly outline the data reduction process here. First, the raw 2-D detector frames have bad pixels identified and masked as well as the instrumental thermal background subtracted. Next, 1-D spectra are extracted assuming a Gaussian line spread function (LSF) and wavelength calibrated. This is done for the companion spectrum and the host star spectrum. For each object, the spectra from each fiber are averaged together with a 3-sigma sigma clipping used to remove outliers, and then the spectra are continuum subtracted using a 100-channel median filter. 

Note that the companion spectrum still contains diffracted light of the star (i.e. speckles) that may be significantly brighter than the flux of the companion. We therefore separate these two components spectrally using the two-component cross correlation described in \citenum{Wang2021}. This two-component cross correlation approach essentially searches for both the companion and speckle spectrum simultaneously by varying the planet's unknown radial velocity. We use the on-axis star spectrum as the stellar spectrum, since any differences should only result in changes in the continuum that should be removed in the continuum subtraction process. For the planet spectrum, we use the same 1400 K atmosphere model from \citenum{Wang2021} containing water and CO opacities and in which the pressure-temperature profile comes from the Sonora-Bobcat model grid\cite{Morley2015}. For each dataset, we obtain a cross correlation function (CCF) which estimates the planet's signal as a function of its radial velocity. A clear peak indicates we have identified a radial velocity for which the companion template matches the signal in the data and constitutes a detection of the companion. Only a subset of the real companions were detected (see Table \ref{tab:obs} for a list of the detected companions). 

\begin{figure}
   \begin{center}
   \begin{tabular}{c} 
   \includegraphics[width=0.75\textwidth]{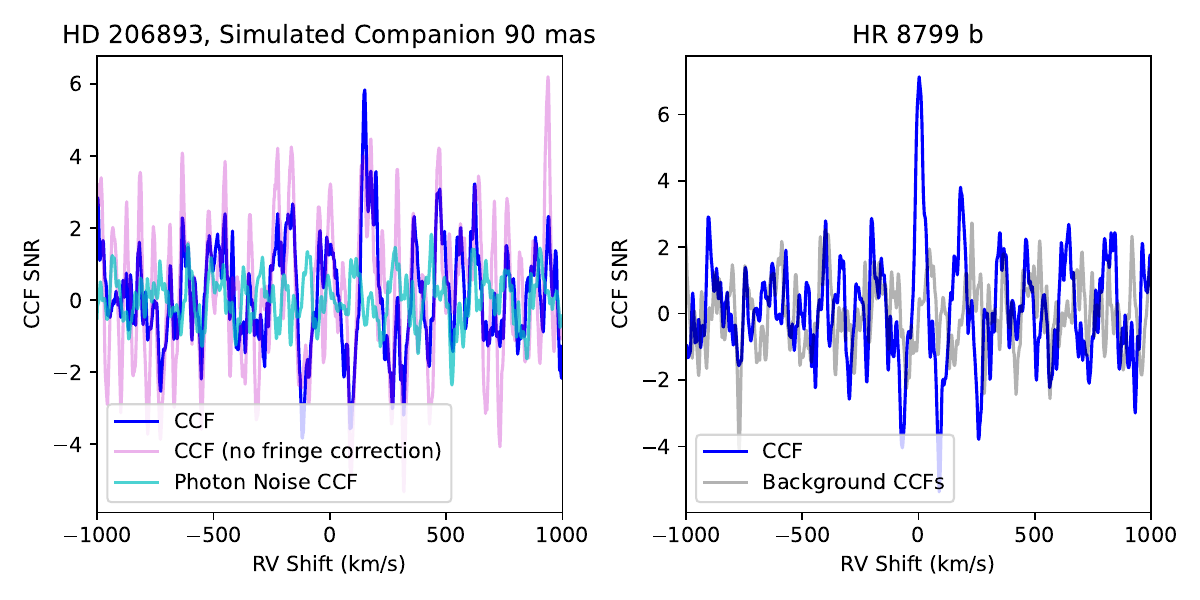}
   \end{tabular}
   \end{center}
   \caption[example] 
   { \label{fig:example} 
Two representative CCFs for a close-in observation with a simulated companion (left) and HR 8799 b, a distant real companion (right). In the left plot, the CCFs of the model template on the data before accounting for fringing (pink) and on a random draw of photon noise (cyan) are also shown. In the right plot, the CCF of the model template with extracted signal of only the NIRSPEC instrument background (gray) is also shown.  }
\end{figure}

For those datasets with non-detections, we injected simulated planetary spectra into the data to determine the sensitivity level. We used a 1450~K, $\log(g) = 4.0$ BT-Settl-CIFIST spectrum\cite{Allard2012} to eumulate the fact that our model template is not a perfect match to any real planetary signal. The spectrum was then convolved with the instrumental LSF, attenuated by the telluric and instrumental response functions, and injected into the non-detection spectrum at a flux ratio that allowed us to visually identify the planet in the CCF. To avoid subtle correlations between any real spectral lines in the data, we injected the simulated spectrum with a radial velocity shift of 150 km/s, which is greater than any planetary orbital velocity. Two example CCFs are shown in Figure \ref{fig:example} (one simulated and one real).

\subsection{Fringing Correction}\label{sec:fringe}
The assumption that the on-axis stellar spectrum can be used to model the speckle spectrum is not perfectly correct due to instrumental distortions. In particular, KPIC spectra suffer from fringing due to multiple plane-parallel transmissive optics along the path \cite{Finnerty2023}. We refer the reader to \citenum{Horstman2024} for details of the fringing, and we will briefly summarize the effect here. The entrance window to the NIRSPEC spectrograph is one of these optics, but it is a static effect that can be calibrated out. However, there are two additional dichroics in the fiber injection unit that also induce fringing. One of them is downstream of the tip-tilt mirror that steers the field so that the light of source of interest is injected into the desired fiber. When switching between the star and the companion, the tip-tilt mirror changes the angle of incidence ($\theta$) onto the dichroic. The modulation ($T$) caused by this fringing follows the equation

\begin{equation}
    T = \left[ 1 + F \sin^2 \left(\frac{2\pi n l \cos\theta}{\lambda}\right) \right]^{-1}
\end{equation}
where $F = 4R/(1-R)^2$, $R$ is the reflectivity of the material, $n$ is the index of refraction of CaF$_2$ as a function of wavelength and temperature, $l$ is the thickness of the material, and $\lambda$ is the wavelength. Thus, changing $\theta$ causes the fringing to change. The fringe frequency due to the dichroic is $\sim 3 \mathring{A}$, and roughly the same frequency as the spectra lines in planetary atmospheres, so it cannot be easily filtered out. Instead, we attempt to model it out in two ways. The simpler method is apply additional fringing to the star data to match the speckle data:
\begin{equation}
    F_{speckle} = F_{star} \times T (\lambda, F, l \cos\theta, T_{CaF2}).
\end{equation}
Here we fit for the reflectance though $F$, the optical delay through the combined term $l \cos\theta$, and the index of refraction through the temperature of the CaF$_2$ glass ($T_{CaF_2}$). This is easier to fit for, but not representative of reality as changing the angle of incidence between the star spectrum and the speckle spectrum does not create an additional plane-parallel optic that induces fringing. However, empirically, this seems to work well.

The alternative method is more physically motivated: the fringing effect of the dichroic changes, so we need to divide the stellar spectrum by the fringing term when we are pointed on-axis on the star ($T_{star}$) and multiply by the fringing term when are are pointed off-axis at the companion ($T_{speckle}$) in order to get the speckle spectrum from the on-axis star spectrum. The equation requires fitting for the on-axis angle of incidence ($\theta_{star}$) and the new angle of incidence ($\theta_{speckle}$):
\begin{equation}
    F_{specke} = F_{star} \times \frac{T (\lambda, F, l \cos\theta_{speckle}, T_{CaF2})}{T (\lambda, F, l \cos\theta_{star}, T_{CaF2})}.
\end{equation}
This optimization is difficult as the terms $\theta_{star}$ and $\theta_{speckle}$ are highly covariant and the 2-D cost function of these two parameters is highly multi-modal. We optimize this using a 2-D grid search, but do not find that it provides a significantly better solution than the previous approximation.

Regardless, we apply either method to all of our observations taken with separations $\leq$ 200 mas to obtain a more accurate speckle spectrum. We rerun the cross correlation on these datasets to obtain a better CCF detection, and use this for assessing our sensitivity. Figure \ref{fig:example} shows one example of the impact of accounting for fringing on the detectability of a simulated companion.

\section{KPIC SENSITIVITY}
\label{sec:sensitivity}

\subsection{Contrast Curve}
\begin{figure}
   \begin{center}
   \begin{tabular}{c} 
   \includegraphics[width=0.8\textwidth]{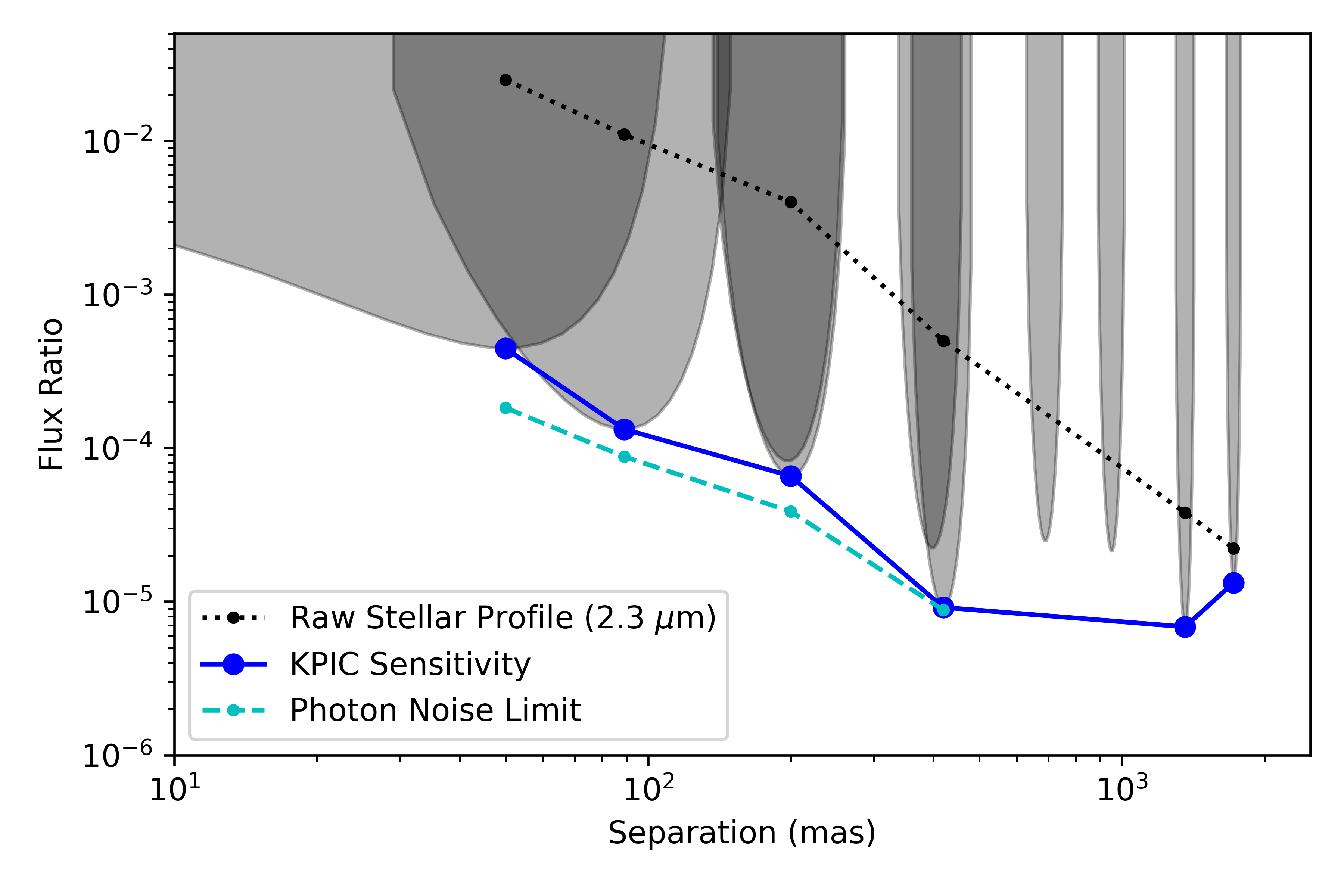}
   \end{tabular}
   \end{center}
   \caption[contrast] 
   { \label{fig:contrast} 
In blue is the adopted contrast curve for KPIC showing its sensitivity in terms of planet-to-star flux ratio as a function of angular separation. Each gray shaded region corresponds to the sensitivity of one dataset used in building up the contrast curve. The dotted black line indicates the total amount of speckle light that leaks into the fiber at each separation. The cyan dashed line shows the photon noise limit at small angular separations. KPIC can reach to within a factor of 2 of the photon noise limit. }
\end{figure}

For each dataset, we can measure the signal-to-noise ratio (SNR) of the detection of either the real or simulated companion in the CCF. With the known $K$-band contrast of the detected real companion or the simulated $K$-band contrast of the simulated companion, we can then estimate the $K$-band flux ratio that would result in a 5-$\sigma$ detection of the companion in the CCF. From empirical testing, we adopt 5-$\sigma$ to be a robust threshold for claiming the detection of a signal. This analysis assumes that we have perfectly positioned our fiber at the desired location. Offsets in the pointing should not be a significant bias, as we have shown that we can offset to companions out to 1500~mas with an accuracy of better than 10~mas (a 10 mas offset corresponds to a $\sim$10\% loss in coupling, which itself should not significantly affect these results). 

In Figure \ref{fig:contrast}, we compile the sensitivity of each dataset as a gray shaded curve. The sensitivities are not delta functions because the fiber does have a finite coupling region, which the amount of flux coupling decreasing as a function of distance from the fiber. We approximate this fiber coupling profile using a perfectly diffraction-limited point spread function created by the Keck II primary mirror to compute the amount of light injected as a function of distance from the fiber. This results in near-parabola like curves for each dataset/pointing (of approximately constant angular width). 

Some datasets were taken around fainter stars, in worse conditions, or earlier in time when instrument performance was not fully optimized. Not including these, we construct a nominal KPIC contrast curve in Figure \ref{fig:contrast} in blue. We also list the measured values in Table \ref{tab:contrast}. For comparison, we also calculate the total amount of stellar speckle light in the companion fiber to show the level of post-processing gain. The level of post-processing gain on average is $\sim$60$\times$. For the separations $\leq$ 400 mas, we also estimate the stellar photon noise floor of our observations.

KPIC is able to reach down to a contrast of $1.3 \times 10^{-4}$ at 90 mas separations from the star. This is within the inner working angles of GPI and SPHERE (see Section \ref{sec:compare}) and reaches planet-level sensitivities (for reference, $\beta$ Pictoris b has a $K$-band flux ratio of $2.1 \times 10^{-4}$) \cite{Bonnefoy2011}. KPIC also achieved a contrast of $9.2 \times 10^{-6}$ at 420 mas from the star. This is two times better than the $K$-band contrast of AF Lep b \cite{DeRosa2023}, and shows that KPIC is able to characterize exoplanets at high contrast. We note that the worsening of contrast at large angular separations beyond 1 arcsec is merely due to reaching the apparent magnitude limit around stars of different brightness (see the following section and Figure \ref{fig:kmag}). 

\begin{table}[ht]
\caption{KPIC Sensitivity} 
\label{tab:contrast}
\begin{center}       
\begin{tabular}{|c|c|c|c|} 
\hline
\rule[1.2ex]{0pt}{3.5ex}  Separation (mas) & Flux Ratio & Kmag & Photon Noise Limit \\
\hline
\rule[-1ex]{0pt}{3.5ex}  50 & $4.4 \times 10^{-4}$ & 14.0 & $1.8 \times 10^{-4}$   \\
\hline
\rule[-1ex]{0pt}{3.5ex} 90 & $1.3 \times 10^{-4}$ & 15.3 & $8.8 \times 10^{-5}$  \\
\hline
\rule[-1ex]{0pt}{3.5ex}  200 & $6.6 \times 10^{-5}$ & 15.7 & $3.9 \times 10^{-5}$  \\
\hline
\rule[-1ex]{0pt}{3.5ex} 420 & $9.2 \times 10^{-6}$ & 17.0 & $8.8 \times 10^{-6}$ \\
\hline
\rule[-1ex]{0pt}{3.5ex} 1359 & $6.9 \times 10^{-6}$ & 17.4 & - \\
\hline 
\rule[-1ex]{0pt}{3.5ex} 1721 & $1.3 \times 10^{-5}$ & 17.4 & -  \\
\hline 
\end{tabular}
\end{center}
\end{table}

\subsection{Apparent Magnitude Sensitivity Curve}
\begin{figure}
   \begin{center}
   \begin{tabular}{c} 
   \includegraphics[width=0.8\textwidth]{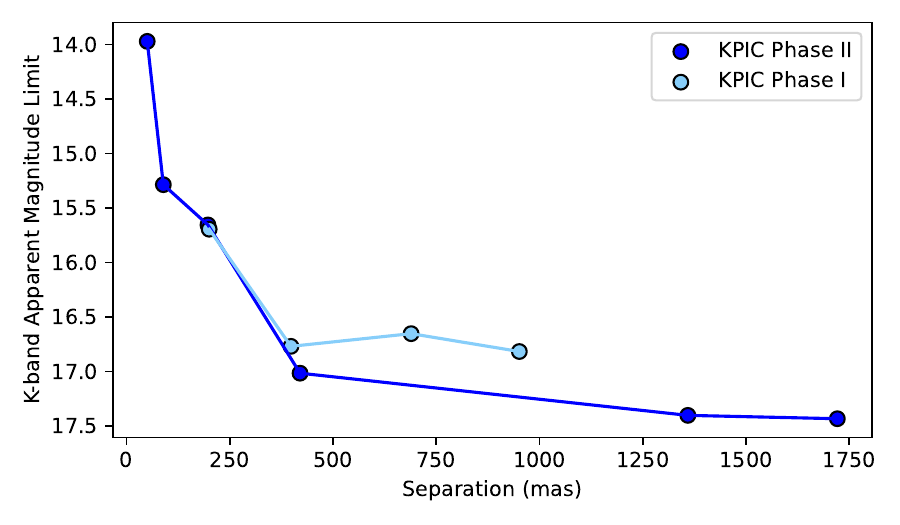}
   \end{tabular}
   \end{center}
   \caption[kmag] 
   { \label{fig:kmag} 
Sensitivity of KPIC in terms of apparent $K$-band magnitude as a function of separation. The performance of KPIC Phase I is plotted in cyan and KPIC Phase II is in blue. KPIC Phase II, as expected, has better sensitivity due to improved throughput. In both cases, at larger separations from the star, the sensitivity levels off. }
\end{figure}

Another way to look at the sensitivity of KPIC is to consider how faint of a companion can be detected by KPIC. We expect that dispersing the light of a planet across tens of thousands of spectral channels would impact how faint of a object we can see, even if the host star is completely suppressed. In Figure \ref{fig:kmag} and Table \ref{tab:contrast}, we show sensitivity curves for KPIC in units of apparent $K$-band magnitude as a function of separation from the star. We separate the performance between KPIC Phase I \cite{Delorme2021} and Phase II \cite{Echeverri2022}, given multiple improvements were made in Phase II to improve the total end-to-end throughput of the instrument -- which leads to a gain in sensitivity of about 0.5 mags. In both cases, we see the curve flatten out at larger separations, indicating we reach some sort of floor in sensitivity, where the amount of starlight leaking into the fiber no longer matters. For KPIC Phase II, this corresponds to an apparent $K$-band magnitude of 17.4. 

To confirm that we are indeed limited by detector noise, we extracted spectra in portions of the detector illuminated by the slit but not illuminated by fiber light (there are four resolution elements between each fiber on the detector). We perform the same cross correlation described in Section \ref{sec:drp} to obtain CCFs of the thermal background. We confirm that the scatter in the thermal background CCFs matches the wings of the CCF where we have a detection. This provides confirmation that the instrument background limits KPIC at the largest separations from the star (e.g., Figure \ref{fig:example}).

\subsection{Comparison with Other High-Contrast Instruments}\label{sec:compare}
\begin{figure}
   \begin{center}
   \begin{tabular}{c} 
   \includegraphics[width=0.8\textwidth]{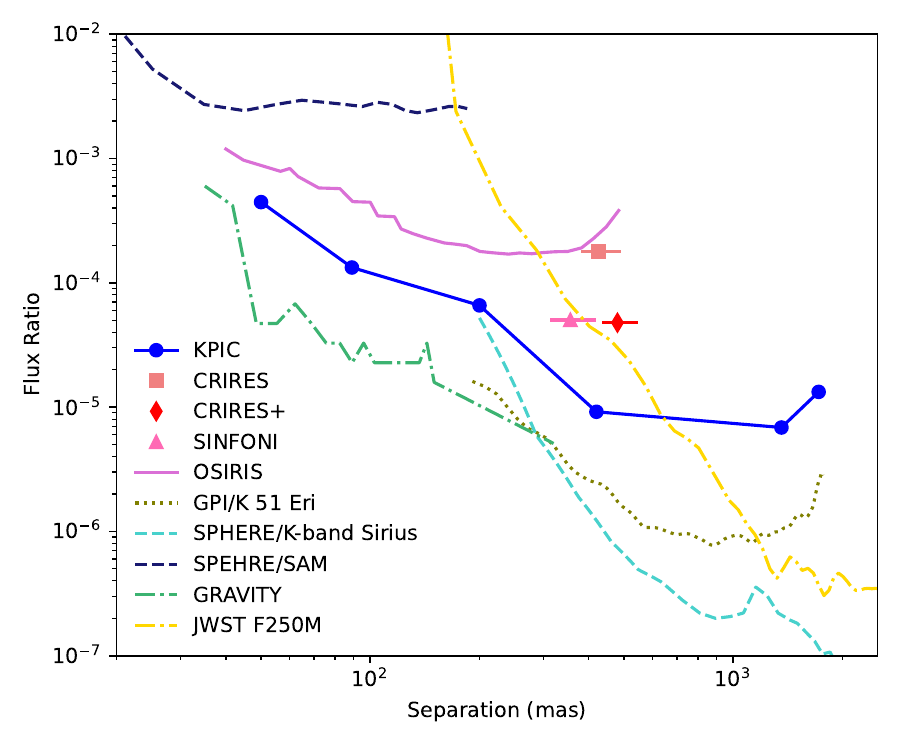}
   \end{tabular}
   \end{center}
   \caption[comparison] 
   { \label{fig:comparison} 
Comparison of the KPIC K-band sensitivity (blue) with other high-contrast imaging instruments. }
\end{figure}

We compare the high-contrast performance of KPIC to other high-contrast instruments operating at $K$ band in Figure \ref{fig:comparison}. For other spectrographs that also use cross correlation to detect companions, we use the CCF SNRs obtained on $\beta$~Pictoris~b from CRIRES\cite{Snellen2014}, SINFONI\cite{Hoeijmakers2018}, and CRIRES+\cite{Landman2024}, along with the CCF-based contrast curve from OSIRIS observations of SR3 \cite{Agrawal2023}. Note that some of these CCF contrast curves are limited to only a single separation. We also include other high-contrast imagers and interferometers. which detect planets in different ways as they do not use the same spectral cross correlation techniques. We include the contrast curves from SPHERE/IRDIS observations of Sirius \cite{Vigan2015}, GPI observations of 51 Eri b \cite{Rajan2017} processed with the Forward Model Matched Filter technique \cite{Ruffio2017}, SPHERE Sparse Aperture Masking (SAM) observations \cite{Stolker2024}, the compilation of VLTI/GRAVITY observations \cite{Pourre2024}, and JWST/NIRCAM observations of HIP 65426 in the F250M filter, which is the closest in wavelength \cite{Carter2023} to our KPIC data. 

There are some caveats when comparing contrast curves derived from different instruments taken with varying integration times and processed with different data analysis methods. Thankfully, the large majority of these observations involve acquisitions spanning over 1 hour in wall-clock time, mitigating the differences in exposure times. We also tried to pick out the best contrast curves available in $K$ band for each instrument. The achieved contrast is also dependent on host-star brightness, so typically brighter contrast can be achieved on brighter stars. In that regard, the SPHERE Sirius observations could be seen as an outlier given the star is orders of magnitude brighter than the rest of the sample. Similarly, the OSIRIS and JWST contrast curves may be pessimistic given that SR3 and HIP 65426 are fainter than the rest of the sample. 

There are also inherent instrument-specific differences. The medium- and high-resolution spectrographs, including KPIC, OSIRIS, SINFONI, CRIRES, and CRIRES+, utilize spectral CCF techniques that discard continuum information for detection purposes. Given that planets are brightest at continuum wavelengths where there are no absorbing molecules, this will inherently limit CCF techniques from reaching the faintest planets. GPI and SPHERE also equipped with adaptive optics systems that can correct more higher-order modes, resulting in less speckle light around the star and more planet photons. 

Out of all of the spectrographs considered that use CCF-based detection techniques, KPIC has the best published high-contrast performance, demonstrating the value of HCI+HRS techniques. We note other spectrographs, such as SINFONI and OSIRIS, should have significantly better sensitivity to companions further out compared to KPIC (e.g., \citenum{Zhang2021} and \citenum{Ruffio2021}), but here we focus on high-contrast capabilities. We notably do not compare against the JWST/NIRSpec high-contrast performance given that it focuses mainly on much longer wavelengths, but the JWST/NIRSpec performance at face value lies between the OSIRIS and KPIC curves \cite{Ruffio2023}.

Also, with the single mode fiber technology, we are able to reach interesting sensitivities ($\sim 10^{-4}$) at separations less than 200 mas, which is inside the inner working angle of GPI, SPHERE, and JWST/NIRCAM. This is corroborated by the performance of GRAVITY at similar separations (50 to 200 mas), which achieves better sensitivity through having 130-meter baselines, separating out the planet in phase, and utilizing the full continuum information of the planet for detection. Both GRAVITY and KPIC demonstrate better high-contrast performance near the diffraction limit than other single-telescope interferometry techniques, such SAM on SPHERE.

\section{LIMITING NOISE SOURCES}
\subsection{High-Contrast Sensitivity Limits}

In this section, we focus on the 50-400 mas separation region, where the contrast curve improves as the fiber moves further from the star. Qualitatively, this means our ability to filter out the light of the speckles becomes the limiting factor. At 400 mas, we successfully reach the photon noise limit, which means that the systematics are not limiting here and that longer exposure times could allow us to go deeper.

Within 400 mas, KPIC does not (yet) reach the photon noise limit. As described in Section \ref{sec:fringe}, fringing is a significant concern, given the fact that the stellar speckle light is orders of magnitude higher than the planet flux. We have tackled fringing in two ways. First, we can use our fringing models described in Section \ref{sec:fringe} to forward model what the effect of fringing should be on the speckle spectrum. Second, in the spring of 2024, we have replaced the KPIC dichroics in the fiber injection unit with wedged versions that should mitigate this effect \cite{Horstman2024}. 

\begin{figure}
   \begin{center}
   \begin{tabular}{c} 
   \includegraphics[width=0.50\textwidth]{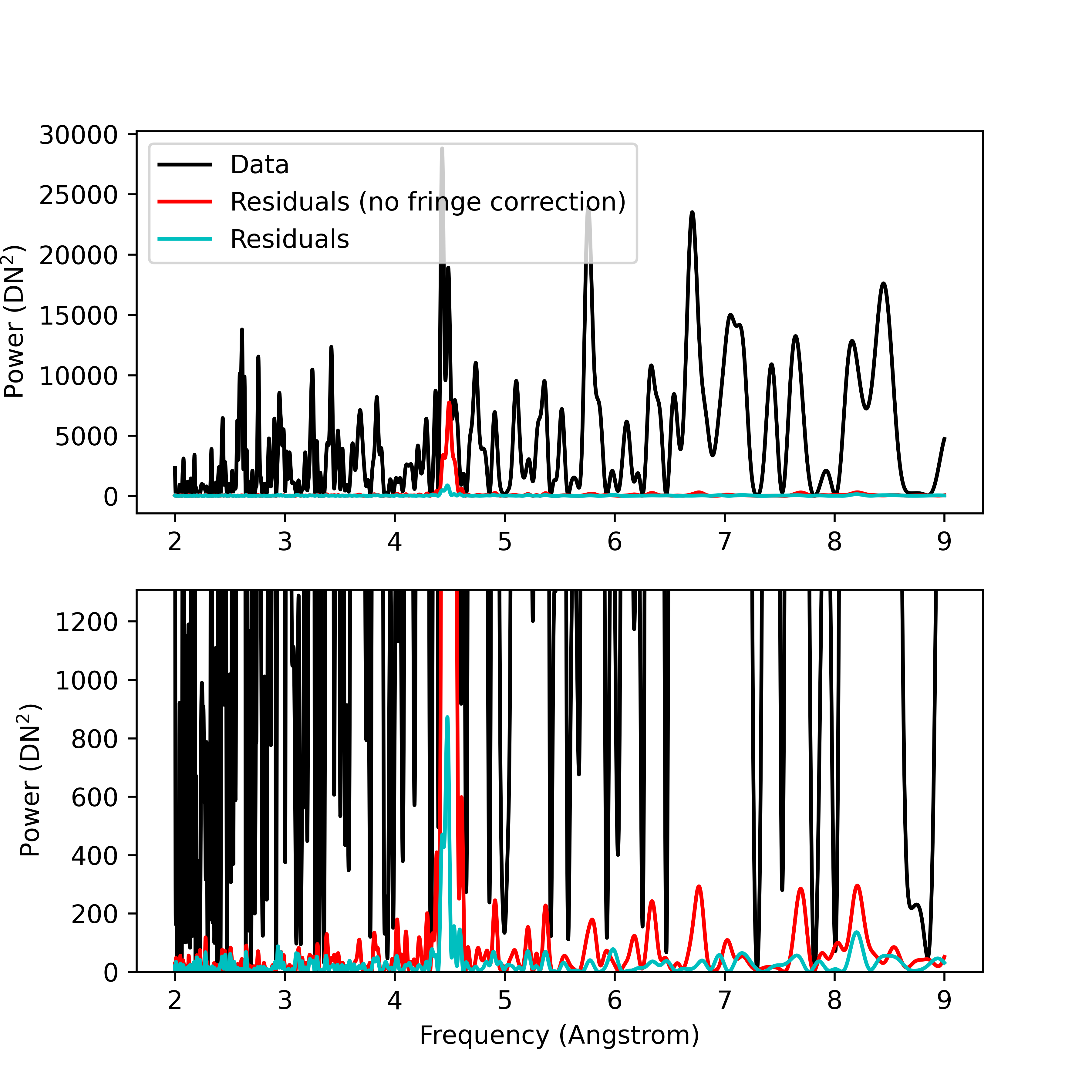}
   \includegraphics[width=0.45\textwidth]{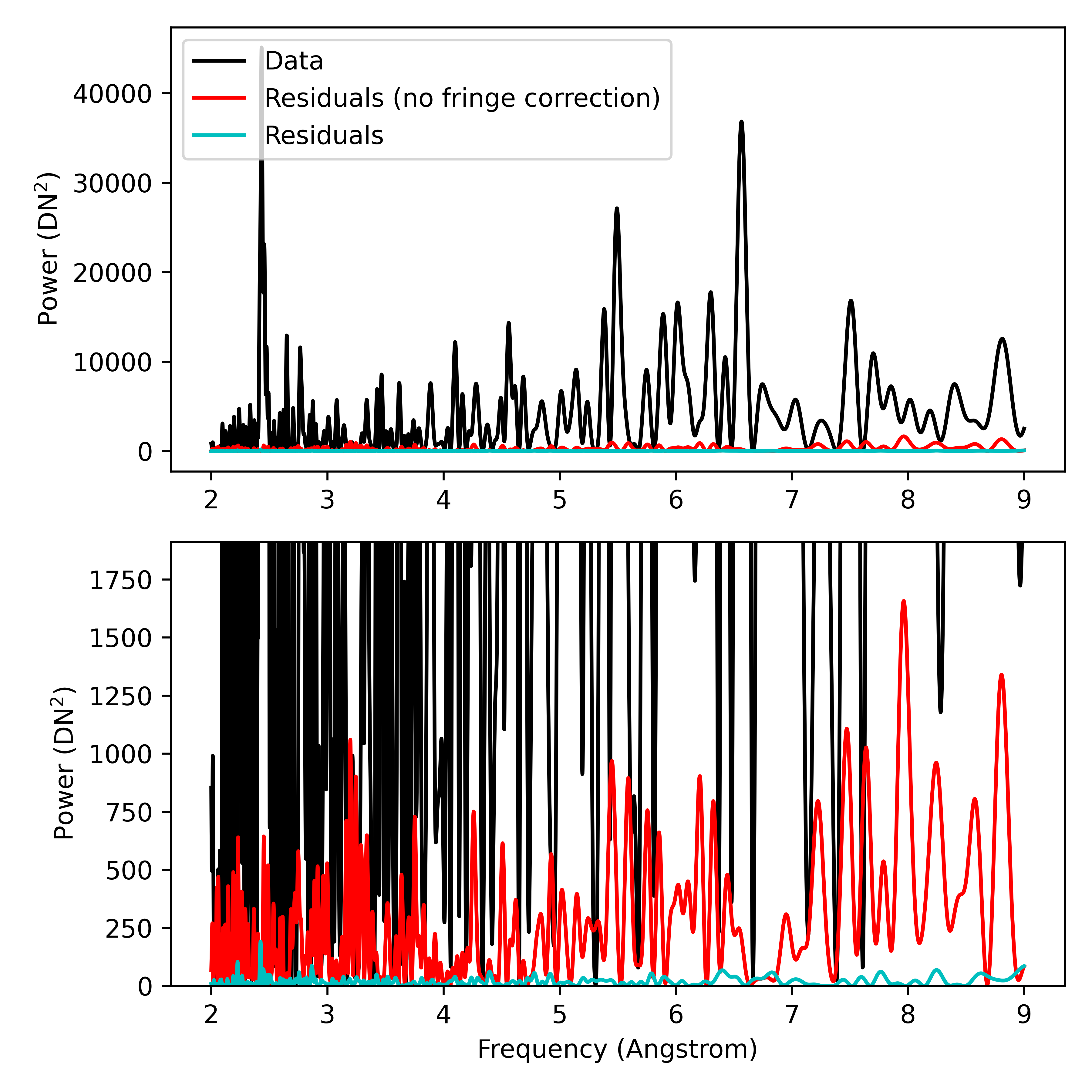}
   \end{tabular}
   \end{center}
   \caption[fringing] 
   { \label{fig:fringing} 
Examples of Lomb-Scargle periodograms before (left column) and after (right column) we replaced the dichroics with wedged versions. In each column, we show the Lomb-Scargle periodogram of the speckle spectrum (black), the residuals after the scaled on-axis stellar spectrum is subtracted from the speckle spectrum (red), and residuals after our model of the fringed spectrum described in Section \ref{sec:fringe} is subtracted from the speckle spectrum (cyan). The bottom panel is a zoomed in version of the top panel to see the scale of the residuals. It is clear that we can substantially model away the fringing, but that wedging the dichroics allows us to almost fully mitigate the effect by itself. }
\end{figure}

We confirm that both methods are effective and even can be used in tandem in Figure \ref{fig:fringing} (see \citenum{Horstman2024} for a more in-depth analysis). With the new dichroics, the fringing signal has almost entirely disappeared, and is difficult to see except in analysis of residuals. Either way, our fringing models can be used to forward model out any residual fringing. With both the wedged dichroics and additional fringe modeling, the fringing signal has been reduced to the level of the noise (bottom right panel of Figure \ref{fig:fringing}). 

With the mitigation of fringing, we are within a factor of 2 of the photon noise limit at separations between 50-200 mas (Figure \ref{fig:contrast}). While fringing is weak enough now that we do not believe it is the limiting factor, there may be other systematics that appear when trying to model starlight that is $\geq$ 50$\times$ brighter than the companion. For example, imperfect modeling due to time-varying telluric features could be an issue given our observations span more than 1 hour typically. A second issue could be systematics in accurately extracting the flux at the 1\% level. If there are systematics in extracting the combined speckle plus planet light, such data reduction artifacts could overwhelm the signal of the planet if the speckle light contribution is very high. Lastly, there may be other unknowns in our complex system that we have not identified, given the number of optics in KPIC. Further work in investigating the source of the remaining systematics is needed. We are very close to the photon noise limit, however, which shows the promise of HCI+HRS to overcome the systematics created by speckles in high contrast imaging systems. 

\subsection{Faint Objective Sensitivity Limits}
\begin{figure}
   \begin{center}
   \begin{tabular}{c} 
   \includegraphics[width=0.45\textwidth]{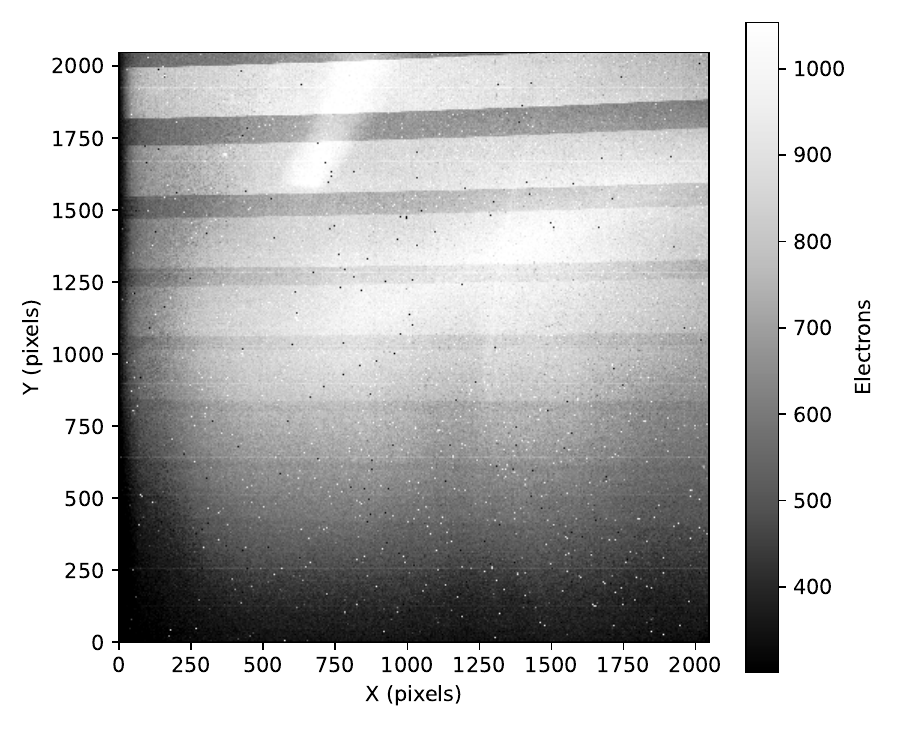}
   \includegraphics[width=0.50\textwidth]{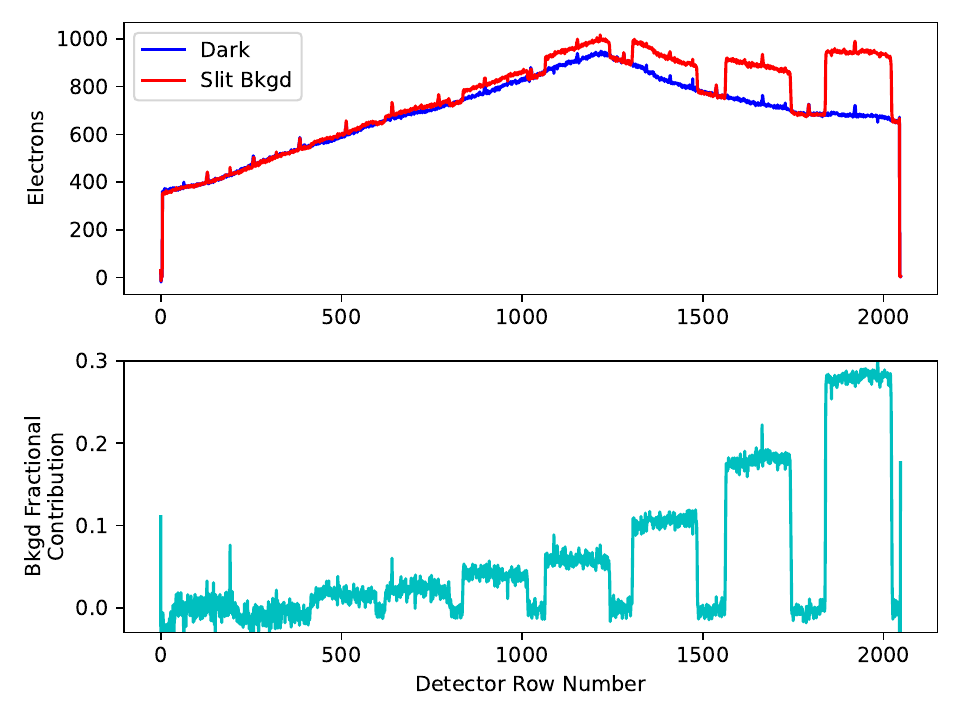}
   \end{tabular}
   \end{center}
   \caption[bkgd] 
   { \label{fig:bkgd} 
\textit{Left:} An example image of the thermal background of KPIC in $K$-band. Nine horizontal echelle orders are seen, with longer wavelengths going to the right and towards the top of the image. In addition to the thermal background through the slit, an elevated dark signal not coming from the slit is seen on the detector \cite{Lopez2020} \textit{Right:} The top plot shows the median flux between columns 1000 and 1100 as a function of row number for both the full thermal background (red) as well as a dark frame where a cold stop is inserted so that no light from the slit is landing on the detector (blue). The bottom plot shows the fraction of the total instrument background that comes from the slit. Even at the longest wavelengths, the majority of the background is due to the elevated dark signal. }
\end{figure}

Given that the noise in the thermal background limits the sensitivity of KPIC, we examine what contributes to the thermal background of the NIRSPEC detector array (previously characterized by \citenum{Martin2018,Lopez2020}). The main difference here is that we use a KPIC-specific pupil stop that is matched to our beam size, so we get a different amount of thermal background coming onto the NIRSPEC detector. We note that the fibers themselves also have a slightly different emissivity than the rest of the KPIC optics, but this difference is very small and negligible for the purposes of considering limiting noise sources. The sources of noise we consider are read noise from the detector, the photon noise due to the elevated dark signal in NIRSPEC, and the photon noise due to the thermal emission of the warm fiber extraction unit (FEU) optics that goes through the slit. 

In Figure \ref{fig:bkgd}, we show an image of a thermal background frame (including both the dark signal from NIRSPEC and the thermal emission through the slit). The read noise is calculated in \citenum{Lopez2020} to be 9.8 electrons, so it will be significantly lower than the 20-30 electrons of photon noise expected given the 400-900 electrons measured in the thermal background frame. When comparing the thermal background frame to one where a cold stop is inserted in the beam to block out all warm thermal emission coming in through the slit, we see in the right panel of Figure \ref{fig:bkgd} that the previously identified elevated dark emission \cite{Lopez2020} is the main source of our instrument background, and thus the photon noise from the elevated dark signal is the main limiting noise source for KPIC. If this can be removed through the identification and masking of possible stray light in the detector, then the background noise in the order covering 2.3 $\mu$m (our best order for making CCF detections; rows 1300-1500 in Figure \ref{fig:bkgd}) will be reduced by a factor of 3, resulting in an improvement of over 1 magnitude in sensitivity. Improvements to the end-to-end through such as higher Strehl ratios, better fiber coupling, or a higher throughput system would also improve the CCF SNR in this detector-noise-limited regime. 

\section{CONCLUSIONS AND FUTURE WORK}
Through a careful assessment of the high-contrast performance of KPIC, we find that the HCI+HRS technique provides improved sensitivity over previously published CCF-based detection approaches. However, KPIC still has not reached the same level of sensitivity as the best datasets from GPI, SPHERE, and GRAVITY, although KPIC can reach sensitivities at the $10^{-4}$-level at 50-200 mas separations, which is within the inner working angle of GPI and SPHERE. At the smallest separations (50-200 mas), KPIC is within a factor of 2 of the photon noise limit after accounting for fringing effects in the spectrum. Future work is needed to identify what remaining sources of systematics limit the high-contrast performance given that the fundamental photon noise limit is within reach. Improvements to the adaptive optics system or new wavefront sensing and control techniques that reduce the amount of stellar speckle light leaking in the fiber should also help with both systematics and the stellar photon noise limit \cite{Xin2023}. For the faintest objects, KPIC is limited by the photon noise due to the elevated dark signal in NIRSPEC and overall system throughput. Mitigating this dark signal or improving the end-to-end efficiency of the system would improve the CCF SNR in this limit. 

The lessons learned from KPIC are applicable to other instruments that utilize HCI+HRS techniques. The HiRISE instrument at VLT \cite{Vigan2024} and the REACH instrument at Subaru \cite{Kotani2020} may encounter similar limiting noise sources, although both use newer spectrographs that may have better performance. The upcoming HISPEC instrument will replace KPIC at Keck and offer significantly better faint object sensitivity due to its better throughput, wider wavelength grasp, and lower instrument background \cite{Mawet2022}. The broader wavelength coverage, better end-to-end throughput, and any future upgrades to the adaptive optics system at Keck will also help with the high-contrast imaging performance of HISPEC, especially if the photon noise limit can be reached. Continued effort in improving stellar speckle spectral calibration is important as reaching the photon noise limit is needed for similar kinds of instruments to search for spectral signatures of habitability in terrestrial exoplanets on the extremely large telescopes \cite{WangJi2017}.

\acknowledgments 
 
Funding for KPIC has been provided by the California Institute of Technology, the Jet Propulsion Laboratory, the Heising-Simons Foundation (grants \#2015-129, \#2017-318, \#2019-1312, \#2023-4598), the Simons Foundation, and the NSF under grant AST-1611623. 
W. M. Keck Observatory access was supported by Northwestern University and the Center for Interdisciplinary Exploration and Research in Astrophysics (CIERA).

The data presented herein were obtained at the W. M. Keck Observatory, which is operated as a scientific partnership among the California Institute of Technology, the University of California and the National Aeronautics and Space Administration. The Observatory was made possible by the generous financial support of the W. M. Keck Foundation. The authors wish to recognize and acknowledge the very significant cultural role and reverence that the summit of Maunakea has always had within the indigenous Hawaiian community.  We are most fortunate to have the opportunity to conduct observations from this mountain. 

\bibliography{report} 

\begin{thebibliography}{10}

\bibitem{Bailey2016}
{Bailey}, V.~P., {Poyneer}, L.~A., {Macintosh}, B.~A., {Savransky}, D., {Wang}, J.~J., {De Rosa}, R.~J., {Follette}, K.~B., {Ammons}, S.~M., {Hayward}, T., {Ingraham}, P., {Maire}, J., {Palmer}, D.~W., {Perrin}, M.~D., {Rajan}, A., {Rantakyr{\"o}}, F.~T., {Thomas}, S., and {V{\'e}ran}, J.-P., ``{Status and performance of the Gemini Planet Imager adaptive optics system},'' in [{\em Adaptive Optics Systems V}{\nolinebreak\hspace{0.1em}]},  {Marchetti}, E., {Close}, L.~M., and {V{\'e}ran}, J.-P., eds., {\em Society of Photo-Optical Instrumentation Engineers (SPIE) Conference Series} {\bf 9909},  99090V (July 2016).

\bibitem{Birkby2013}
{Birkby}, J.~L., {de Kok}, R.~J., {Brogi}, M., {de Mooij}, E.~J.~W., {Schwarz}, H., {Albrecht}, S., and {Snellen}, I.~A.~G., ``{Detection of water absorption in the day side atmosphere of HD 189733 b using ground-based high-resolution spectroscopy at 3.2{\ensuremath{\mu}}m.},'' {\em MNRAS}~{\bf 436},  L35--L39 (Nov. 2013).

\bibitem{Line2021}
{Line}, M.~R., {Brogi}, M., {Bean}, J.~L., {Gandhi}, S., {Zalesky}, J., {Parmentier}, V., {Smith}, P., {Mace}, G.~N., {Mansfield}, M., {Kempton}, E. M.~R., {Fortney}, J.~J., {Shkolnik}, E., {Patience}, J., {Rauscher}, E., {D{\'e}sert}, J.-M., and {Wardenier}, J.~P., ``{A solar C/O and sub-solar metallicity in a hot Jupiter atmosphere},'' {\em Nature}~{\bf 598},  580--584 (Oct. 2021).

\bibitem{Finnerty2023}
{Finnerty}, L., {Horstman}, K., {Ruffio}, J.-B., {Wang}, J.~J., {Mawet}, D., {Schofield}, T., {Sappey}, B., {Xuan}, J., {Delorme}, J.-R., {Jovanovic}, N., {Blake}, G.~A., and {Fitzgerald}, M.~P., ``{Characterization of hot Jupiter atmospheres with Keck/KPIC},'' in [{\em Society of Photo-Optical Instrumentation Engineers (SPIE) Conference Series}{\nolinebreak\hspace{0.1em}]},  {\em Society of Photo-Optical Instrumentation Engineers (SPIE) Conference Series} {\bf 12680},  1268006 (Oct. 2023).

\bibitem{Snellen2014}
{Snellen}, I. A.~G., {Brandl}, B.~R., {de Kok}, R.~J., {Brogi}, M., {Birkby}, J., and {Schwarz}, H., ``{Fast spin of the young extrasolar planet {\ensuremath{\beta}} Pictoris b},'' {\em Nature}~{\bf 509},  63--65 (May 2014).

\bibitem{Hoeijmakers2018}
{Hoeijmakers}, H.~J., {Schwarz}, H., {Snellen}, I.~A.~G., {de Kok}, R.~J., {Bonnefoy}, M., {Chauvin}, G., {Lagrange}, A.~M., and {Girard}, J.~H., ``{Medium-resolution integral-field spectroscopy for high-contrast exoplanet imaging. Molecule maps of the {\ensuremath{\beta}} Pictoris system with SINFONI},'' {\em A\&A}~{\bf 617},  A144 (Oct. 2018).

\bibitem{Agrawal2023}
{Agrawal}, S., {Ruffio}, J.-B., {Konopacky}, Q.~M., {Macintosh}, B., {Mawet}, D., {Nielsen}, E.~L., {Hoch}, K. K.~W., {Liu}, M.~C., {Barman}, T.~S., {Thompson}, W., {Greenbaum}, A.~Z., {Marois}, C., and {Patience}, J., ``{Detecting Exoplanets Closer to Stars with Moderate Spectral Resolution Integral-field Spectroscopy},'' {\em AJ}~{\bf 166},  15 (July 2023).

\bibitem{Kasper2021}
{Kasper}, M., {Cerpa Urra}, N., {Pathak}, P., {Bonse}, M., {Nousiainen}, J., {Engler}, B., {Heritier}, C.~T., {Kammerer}, J., {Leveratto}, S., {Rajani}, C., {Bristow}, P., {Le Louarn}, M., {Madec}, P.~Y., {Str{\"o}bele}, S., {Verinaud}, C., {Glauser}, A., {Quanz}, S.~P., {Helin}, T., {Keller}, C., {Snik}, F., {Boccaletti}, A., {Chauvin}, G., {Mouillet}, D., {Kulcs{\'a}r}, C., and {Raynaud}, H.~F., ``{PCS {\textemdash} A Roadmap for Exoearth Imaging with the ELT},'' {\em The Messenger}~{\bf 182},  38--43 (Mar. 2021).

\bibitem{Mawet2022}
{Mawet}, D., {Fitzgerald}, M.~P., {Konopacky}, Q., {Jovanovic}, N., {Baker}, A., {Beichman}, C., {Bertz}, R., {Dekany}, R., {Fucik}, J., {Roberts}, M., {Porter}, M., {Pahuja}, R., {Ruane}, G., {Leifer}, S., {Halverson}, S., {Gibbs}, A., {Johnson}, C., {Kress}, E., {Magnone}, K., {Sohn}, J.~M., {Wang}, E., {Brown}, A., {Maire}, J., {Sappey}, B., {Andersen}, D., {Terada}, H., {Kassis}, M., {Artigau}, E., {Benneke}, B., {Doyon}, R., {Kotani}, T., {Tamura}, M., {Beatty}, T., {Plavchan}, P., {Do}, T., {Nishiyama}, S., {Wang}, J., and {Wang}, J., ``{Fiber-fed high-resolution infrared spectroscopy at the diffraction limit with Keck-HISPEC and TMT-MODHIS: status update},'' in [{\em Ground-based and Airborne Instrumentation for Astronomy IX}{\nolinebreak\hspace{0.1em}]},  {Evans}, C.~J., {Bryant}, J.~J., and {Motohara}, K., eds., {\em Society of Photo-Optical Instrumentation Engineers (SPIE) Conference Series} {\bf 12184},  121841R (Aug. 2022).

\bibitem{Kautz2023}
{Kautz}, M., {Males}, J.~R., {Close}, L.~M., {Haffert}, S.~Y., {Guyon}, O., {Hedglen}, A., {Gasho}, V., {Durney}, O., {Noenickx}, J., {Fletcher}, A., {Coronado}, F., {Ford}, J., {Connors}, T., {Sullivan}, M., {Salanski}, T., {Kelly}, D., {Demers}, R., {Bouchez}, A., {Sitarski}, B., and {Schurter}, P., ``{GMagAO-X: A First Light Coronagraphic Adaptive Optics System for the GMT},'' {\em arXiv e-prints} ,  arXiv:2310.10888 (Oct. 2023).

\bibitem{WangJi2017}
{Wang}, J., {Mawet}, D., {Ruane}, G., {Hu}, R., and {Benneke}, B., ``{Observing Exoplanets with High Dispersion Coronagraphy. I. The Scientific Potential of Current and Next-generation Large Ground and Space Telescopes},'' {\em AJ}~{\bf 153},  183 (Apr. 2017).

\bibitem{Mawet2017}
{Mawet}, D., {Delorme}, J.~R., {Jovanovic}, N., {Wallace}, J.~K., {Bartos}, R.~D., {Wizinowich}, P.~L., {Fitzgerald}, M., {Lilley}, S., {Ruane}, G., {Wang}, J., {Klimovich}, N., and {Xin}, Y., ``{A fiber injection unit for the Keck Planet Imager and Characterizer},'' in [{\em Society of Photo-Optical Instrumentation Engineers (SPIE) Conference Series}{\nolinebreak\hspace{0.1em}]},  {\em Society of Photo-Optical Instrumentation Engineers (SPIE) Conference Series} {\bf 10400},  1040029 (Sept. 2017).

\bibitem{Delorme2021}
{Delorme}, J.-R., {Jovanovic}, N., {Echeverri}, D., {Mawet}, D., {Kent Wallace}, J., {Bartos}, R.~D., {Cetre}, S., {Wizinowich}, P., {Ragland}, S., {Lilley}, S., {Wetherell}, E., {Doppmann}, G., {Wang}, J.~J., {Morris}, E.~C., {Ruffio}, J.-B., {Martin}, E.~C., {Fitzgerald}, M.~P., {Ruane}, G., {Schofield}, T., {Suominen}, N., {Calvin}, B., {Wang}, E., {Magnone}, K., {Johnson}, C., {Sohn}, J.~M., {L{\'o}pez}, R.~A., {Bond}, C.~Z., {Pezzato}, J., {Sayson}, J.~L., {Chun}, M., and {Skemer}, A.~J., ``{Keck Planet Imager and Characterizer: a dedicated single-mode fiber injection unit for high-resolution exoplanet spectroscopy},'' {\em Journal of Astronomical Telescopes, Instruments, and Systems}~{\bf 7},  035006 (July 2021).

\bibitem{Echeverri2022}
{Echeverri}, D., {Jovanovic}, N., {Delorme}, J.-R., {Xin}, Y., {Schofield}, T., {Finnerty}, L., {Wang}, J.~J., {Xuan}, J., {Mawet}, D., {Baker}, A., {Bartos}, R., {Bond}, C.~Z., {Bryan}, M., {Calvin}, B., {Cetre}, S., {Doppmann}, G., {Fitzgerald}, M.~P., {Fucik}, J., {Horstman}, K., {Lopez}, R., {Martin}, E.~C., {Martin}, S., {Mennesson}, B., {Morris}, E., {Nash}, R., {Pezzato}, J., {Porter}, M., {Ragland}, S., {Roberts}, M.~K., {Ruane}, G., {Ruffio}, J.-B., {Sappey}, B., {Serabyn}, E., {Skemer}, A., {Venenciano}, T., {Wallace}, J.~K., {Wang}, J., and {Wizinowich}, P., ``{Phase II of the Keck Planet Imager and characterizer: system-level laboratory characterization and preliminary on-sky commissioning},'' in [{\em Ground-based and Airborne Instrumentation for Astronomy IX}{\nolinebreak\hspace{0.1em}]},  {Evans}, C.~J., {Bryant}, J.~J., and {Motohara}, K., eds., {\em Society of Photo-Optical Instrumentation Engineers (SPIE) Conference Series} {\bf 12184},  121841W (Aug. 2022).

\bibitem{McLean1998}
{McLean}, I.~S., {Becklin}, E.~E., {Bendiksen}, O., {Brims}, G., {Canfield}, J., {Figer}, D.~F., {Graham}, J.~R., {Hare}, J., {Lacayanga}, F., {Larkin}, J.~E., {Larson}, S.~B., {Levenson}, N., {Magnone}, N., {Teplitz}, H., and {Wong}, W., ``{Design and development of NIRSPEC: a near-infrared echelle spectrograph for the Keck II telescope},'' in [{\em Infrared Astronomical Instrumentation}{\nolinebreak\hspace{0.1em}]},  {Fowler}, A.~M., ed., {\em Society of Photo-Optical Instrumentation Engineers (SPIE) Conference Series} {\bf 3354},  566--578 (Aug. 1998).

\bibitem{Martin2018}
{Martin}, E.~C., {Fitzgerald}, M.~P., {McLean}, I.~S., {Doppmann}, G., {Kassis}, M., {Aliado}, T., {Canfield}, J., {Johnson}, C., {Kress}, E., {Lanclos}, K., {Magnone}, K., {Sohn}, J.~M., {Wang}, E., and {Weiss}, J., ``{An overview of the NIRSPEC upgrade for the Keck II telescope},'' in [{\em Ground-based and Airborne Instrumentation for Astronomy VII}{\nolinebreak\hspace{0.1em}]},  {Evans}, C.~J., {Simard}, L., and {Takami}, H., eds., {\em Society of Photo-Optical Instrumentation Engineers (SPIE) Conference Series} {\bf 10702},  107020A (Aug. 2018).

\bibitem{Lopez2020}
{L{\'o}pez}, R.~A., {Hoffman}, E.~B., {Doppmann}, G., {Fitzgerald}, M.~P., {Johnson}, C., {Kassis}, M., {Lanclos}, K., {Lyke}, J., {Martin}, E.~C., {McLean}, I., {Sohn}, J.~M., and {Weiss}, J., ``{Characterization and performance of the upgraded NIRSPEC on the W. M. Keck Telescope},'' in [{\em Ground-based and Airborne Instrumentation for Astronomy VIII}{\nolinebreak\hspace{0.1em}]},  {Evans}, C.~J., {Bryant}, J.~J., and {Motohara}, K., eds., {\em Society of Photo-Optical Instrumentation Engineers (SPIE) Conference Series} {\bf 11447},  114476B (Dec. 2020).

\bibitem{Wang2021}
{Wang}, J.~J., {Ruffio}, J.-B., {Morris}, E., {Delorme}, J.-R., {Jovanovic}, N., {Pezzato}, J., {Echeverri}, D., {Finnerty}, L., {Hood}, C., {Zanazzi}, J.~J., {Bryan}, M.~L., {Bond}, C.~Z., {Cetre}, S., {Martin}, E.~C., {Mawet}, D., {Skemer}, A., {Baker}, A., {Xuan}, J.~W., {Wallace}, J.~K., {Wang}, J., {Bartos}, R., {Blake}, G.~A., {Boden}, A., {Buzard}, C., {Calvin}, B., {Chun}, M., {Doppmann}, G., {Dupuy}, T.~J., {Duch{\^e}ne}, G., {Feng}, Y.~K., {Fitzgerald}, M.~P., {Fortney}, J., {Freedman}, R.~S., {Knutson}, H., {Konopacky}, Q., {Lilley}, S., {Liu}, M.~C., {Lopez}, R., {Lupu}, R., {Marley}, M.~S., {Meshkat}, T., {Miles}, B., {Millar-Blanchaer}, M., {Ragland}, S., {Roy}, A., {Ruane}, G., {Sappey}, B., {Schofield}, T., {Weiss}, L., {Wetherell}, E., {Wizinowich}, P., and {Ygouf}, M., ``{Detection and Bulk Properties of the HR 8799 Planets with High-resolution Spectroscopy},'' {\em AJ}~{\bf 162},  148 (Oct. 2021).

\bibitem{Xin2023}
{Xin}, Y., {Xuan}, J.~W., {Mawet}, D., {Wang}, J., {Ruane}, G., {Echeverri}, D., {Jovanovic}, N., {Do {\~A}`}, C., {Fitzgerald}, M., {Horstman}, K., {Hsu}, C.-C., {Liberman}, J., {L{\'o}pez}, R.~A., {Phillips}, C.~L., {Ren}, B.~B., {Ruffio}, J.-B., and {Sappey}, B., ``{On-sky speckle nulling through a single-mode fiber with the Keck Planet Imager and Characterizer},'' {\em Journal of Astronomical Telescopes, Instruments, and Systems}~{\bf 9},  035001 (July 2023).

\bibitem{Cutri2003}
{Cutri}, R.~M., {Skrutskie}, M.~F., {van Dyk}, S., {Beichman}, C.~A., {Carpenter}, J.~M., {Chester}, T., {Cambresy}, L., {Evans}, T., {Fowler}, J., {Gizis}, J., {Howard}, E., {Huchra}, J., {Jarrett}, T., {Kopan}, E.~L., {Kirkpatrick}, J.~D., {Light}, R.~M., {Marsh}, K.~A., {McCallon}, H., {Schneider}, S., {Stiening}, R., {Sykes}, M., {Weinberg}, M., {Wheaton}, W.~A., {Wheelock}, S., and {Zacarias}, N., ``{VizieR Online Data Catalog: 2MASS All-Sky Catalog of Point Sources (Cutri+ 2003)}.'' VizieR On-line Data Catalog: II/246. Originally published in: 2003yCat.2246....0C (June 2003).

\bibitem{witp}
{Wang}, J.~J., {Kulikauskas}, M., and {Blunt}, S., ``{whereistheplanet: Predicting positions of directly imaged companions}.'' Astrophysics Source Code Library, record ascl:2101.003 (Jan. 2021).

\bibitem{Zurlo2016}
{Zurlo}, A., {Vigan}, A., {Galicher}, R., {Maire}, A.~L., {Mesa}, D., {Gratton}, R., {Chauvin}, G., {Kasper}, M., {Moutou}, C., {Bonnefoy}, M., {Desidera}, S., {Abe}, L., {Apai}, D., {Baruffolo}, A., {Baudoz}, P., {Baudrand}, J., {Beuzit}, J.~L., {Blancard}, P., {Boccaletti}, A., {Cantalloube}, F., {Carle}, M., {Cascone}, E., {Charton}, J., {Claudi}, R.~U., {Costille}, A., {de Caprio}, V., {Dohlen}, K., {Dominik}, C., {Fantinel}, D., {Feautrier}, P., {Feldt}, M., {Fusco}, T., {Gigan}, P., {Girard}, J.~H., {Gisler}, D., {Gluck}, L., {Gry}, C., {Henning}, T., {Hugot}, E., {Janson}, M., {Jaquet}, M., {Lagrange}, A.~M., {Langlois}, M., {Llored}, M., {Madec}, F., {Magnard}, Y., {Martinez}, P., {Maurel}, D., {Mawet}, D., {Meyer}, M.~R., {Milli}, J., {Moeller-Nilsson}, O., {Mouillet}, D., {Orign{\'e}}, A., {Pavlov}, A., {Petit}, C., {Puget}, P., {Quanz}, S.~P., {Rabou}, P., {Ramos}, J., {Rousset}, G., {Roux}, A., {Salasnich}, B., {Salter}, G., {Sauvage}, J.~F., {Schmid}, H.~M., {Soenke}, C., {Stadler}, E., {Suarez},
  M., {Turatto}, M., {Udry}, S., {Vakili}, F., {Wahhaj}, Z., {Wildi}, F., and {Antichi}, J., ``{First light of the VLT planet finder SPHERE. III. New spectrophotometry and astrometry of the HR 8799 exoplanetary system},'' {\em A\&A}~{\bf 587},  A57 (Mar. 2016).

\bibitem{Delorme2017}
{Delorme}, P., {Schmidt}, T., {Bonnefoy}, M., {Desidera}, S., {Ginski}, C., {Charnay}, B., {Lazzoni}, C., {Christiaens}, V., {Messina}, S., {D'Orazi}, V., {Milli}, J., {Schlieder}, J.~E., {Gratton}, R., {Rodet}, L., {Lagrange}, A.~M., {Absil}, O., {Vigan}, A., {Galicher}, R., {Hagelberg}, J., {Bonavita}, M., {Lavie}, B., {Zurlo}, A., {Olofsson}, J., {Boccaletti}, A., {Cantalloube}, F., {Mouillet}, D., {Chauvin}, G., {Hambsch}, F.~J., {Langlois}, M., {Udry}, S., {Henning}, T., {Beuzit}, J.~L., {Mordasini}, C., {Lucas}, P., {Marocco}, F., {Biller}, B., {Carson}, J., {Cheetham}, A., {Covino}, E., {De Caprio}, V., {Delboulbe}, A., {Feldt}, M., {Girard}, J., {Hubin}, N., {Maire}, A.~L., {Pavlov}, A., {Petit}, C., {Rouan}, D., {Roelfsema}, R., and {Wildi}, F., ``{In-depth study of moderately young but extremely red, very dusty substellar companion HD 206893B},'' {\em A\&A}~{\bf 608},  A79 (Dec. 2017).

\bibitem{Hinkley2023}
{Hinkley}, S., {Lacour}, S., {Marleau}, G.~D., {Lagrange}, A.~M., {Wang}, J.~J., {Kammerer}, J., {Cumming}, A., {Nowak}, M., {Rodet}, L., {Stolker}, T., {Balmer}, W.~O., {Ray}, S., {Bonnefoy}, M., {Molli{\`e}re}, P., {Lazzoni}, C., {Kennedy}, G., {Mordasini}, C., {Abuter}, R., {Aigrain}, S., {Amorim}, A., {Asensio-Torres}, R., {Babusiaux}, C., {Benisty}, M., {Berger}, J.~P., {Beust}, H., {Blunt}, S., {Boccaletti}, A., {Bohn}, A., {Bonnet}, H., {Bourdarot}, G., {Brandner}, W., {Cantalloube}, F., {Caselli}, P., {Charnay}, B., {Chauvin}, G., {Chomez}, A., {Choquet}, E., {Christiaens}, V., {Cl{\'e}net}, Y., {Coud{\'e} du Foresto}, V., {Cridland}, A., {Delorme}, P., {Dembet}, R., {Drescher}, A., {Duvert}, G., {Eckart}, A., {Eisenhauer}, F., {Feuchtgruber}, H., {Galland}, F., {Garcia}, P., {Garcia Lopez}, R., {Gardner}, T., {Gendron}, E., {Genzel}, R., {Gillessen}, S., {Girard}, J.~H., {Grandjean}, A., {Haubois}, X., {Hei{\ss}el}, G., {Henning}, T., {Hippler}, S., {Horrobin}, M., {Houll{\'e}}, M., {Hubert}, Z.,
  {Jocou}, L., {Keppler}, M., {Kervella}, P., {Kreidberg}, L., {Lapeyr{\`e}re}, V., {Le Bouquin}, J.~B., {L{\'e}na}, P., {Lutz}, D., {Maire}, A.~L., {Mang}, F., {M{\'e}rand}, A., {Meunier}, N., {Monnier}, J.~D., {Mouillet}, D., {Nasedkin}, E., {Ott}, T., {Otten}, G.~P.~P.~L., {Paladini}, C., {Paumard}, T., {Perraut}, K., {Perrin}, G., {Philipot}, F., {Pfuhl}, O., {Pourr{\'e}}, N., {Pueyo}, L., {Rameau}, J., {Rickman}, E., {Rubini}, P., {Rustamkulov}, Z., {Samland}, M., {Shangguan}, J., {Shimizu}, T., {Sing}, D., {Straubmeier}, C., {Sturm}, E., {Tacconi}, L.~J., {van Dishoeck}, E.~F., {Vigan}, A., {Vincent}, F., {Ward-Duong}, K., {Widmann}, F., {Wieprecht}, E., {Wiezorrek}, E., {Woillez}, J., {Yazici}, S., {Young}, A., and {Zicher}, N., ``{Direct discovery of the inner exoplanet in the HD 206893 system. Evidence for deuterium burning in a planetary-mass companion},'' {\em A\&A}~{\bf 671},  L5 (Mar. 2023).

\bibitem{Currie2023}
{Currie}, T., {Brandt}, G.~M., {Brandt}, T.~D., {Lacy}, B., {Burrows}, A., {Guyon}, O., {Tamura}, M., {Liu}, R.~Y., {Sagynbayeva}, S., {Tobin}, T., {Chilcote}, J., {Groff}, T., {Marois}, C., {Thompson}, W., {Murphy}, S.~J., {Kuzuhara}, M., {Lawson}, K., {Lozi}, J., {Deo}, V., {Vievard}, S., {Skaf}, N., {Uyama}, T., {Jovanovic}, N., {Martinache}, F., {Kasdin}, N.~J., {Kudo}, T., {McElwain}, M., {Janson}, M., {Wisniewski}, J., {Hodapp}, K., {Nishikawa}, J., {He{\l}miniak}, K., {Kwon}, J., and {Hayashi}, M., ``{Direct imaging and astrometric detection of a gas giant planet orbiting an accelerating star},'' {\em Science}~{\bf 380},  198--203 (Apr. 2023).

\bibitem{Vigan2016}
{Vigan}, A., {Bonnefoy}, M., {Ginski}, C., {Beust}, H., {Galicher}, R., {Janson}, M., {Baudino}, J.~L., {Buenzli}, E., {Hagelberg}, J., {D'Orazi}, V., {Desidera}, S., {Maire}, A.~L., {Gratton}, R., {Sauvage}, J.~F., {Chauvin}, G., {Thalmann}, C., {Malo}, L., {Salter}, G., {Zurlo}, A., {Antichi}, J., {Baruffolo}, A., {Baudoz}, P., {Blanchard}, P., {Boccaletti}, A., {Beuzit}, J.~L., {Carle}, M., {Claudi}, R., {Costille}, A., {Delboulb{\'e}}, A., {Dohlen}, K., {Dominik}, C., {Feldt}, M., {Fusco}, T., {Gluck}, L., {Girard}, J., {Giro}, E., {Gry}, C., {Henning}, T., {Hubin}, N., {Hugot}, E., {Jaquet}, M., {Kasper}, M., {Lagrange}, A.~M., {Langlois}, M., {Le Mignant}, D., {Llored}, M., {Madec}, F., {Martinez}, P., {Mawet}, D., {Mesa}, D., {Milli}, J., {Mouillet}, D., {Moulin}, T., {Moutou}, C., {Orign{\'e}}, A., {Pavlov}, A., {Perret}, D., {Petit}, C., {Pragt}, J., {Puget}, P., {Rabou}, P., {Rochat}, S., {Roelfsema}, R., {Salasnich}, B., {Schmid}, H.~M., {Sevin}, A., {Siebenmorgen}, R., {Smette}, A., {Stadler}, E.,
  {Suarez}, M., {Turatto}, M., {Udry}, S., {Vakili}, F., {Wahhaj}, Z., {Weber}, L., and {Wildi}, F., ``{First light of the VLT planet finder SPHERE. I. Detection and characterization of the substellar companion GJ 758 B},'' {\em A\&A}~{\bf 587},  A55 (Mar. 2016).

\bibitem{Morley2015}
{Morley}, C.~V., {Fortney}, J.~J., {Marley}, M.~S., {Zahnle}, K., {Line}, M., {Kempton}, E., {Lewis}, N., and {Cahoy}, K., ``{Thermal Emission and Reflected Light Spectra of Super Earths with Flat Transmission Spectra},'' {\em ApJ}~{\bf 815},  110 (Dec. 2015).

\bibitem{Allard2012}
{Allard}, F., {Homeier}, D., and {Freytag}, B., ``{Models of very-low-mass stars, brown dwarfs and exoplanets},'' {\em Philosophical Transactions of the Royal Society of London Series A}~{\bf 370},  2765--2777 (Jun 2012).

\bibitem{Horstman2024}
Horstman, K., Ruffio, J.-B., Wang, J.~J., Hsu, C.-C., Baker, A., Finnerty, L., Xuan, J.~W., Echeverri, D., Mawet, D., and Team, K., ``Fringing analysis and forward modeling of keck planet imager and characterizer (kpic) spectra,'' in [{\em This Conference}{\nolinebreak\hspace{0.1em}]},  (2024).

\bibitem{Bonnefoy2011}
{Bonnefoy}, M., {Lagrange}, A.~M., {Boccaletti}, A., {Chauvin}, G., {Apai}, D., {Allard}, F., {Ehrenreich}, D., {Girard}, J.~H.~V., {Mouillet}, D., {Rouan}, D., {Gratadour}, D., and {Kasper}, M., ``{High angular resolution detection of {\ensuremath{\beta}} Pictoris b at 2.18 {\ensuremath{\mu}}m},'' {\em A\&A}~{\bf 528},  L15 (Apr. 2011).

\bibitem{DeRosa2023}
{De Rosa}, R.~J., {Nielsen}, E.~L., {Wahhaj}, Z., {Ruffio}, J.-B., {Kalas}, P.~G., {Peck}, A.~E., {Hirsch}, L.~A., and {Roberson}, W., ``{Direct imaging discovery of a super-Jovian around the young Sun-like star AF Leporis},'' {\em A\&A}~{\bf 672},  A94 (Apr. 2023).

\bibitem{Landman2024}
{Landman}, R., {Stolker}, T., {Snellen}, I.~A.~G., {Costes}, J., {de Regt}, S., {Zhang}, Y., {Gandhi}, S., {Molliere}, P., {Kesseli}, A., {Vigan}, A., and {Sanchez-L{\'o}pez}, A., ``{{\ensuremath{\beta}} Pictoris b through the eyes of the upgraded CRIRES+. Atmospheric composition, spin rotation, and radial velocity},'' {\em A\&A}~{\bf 682},  A48 (Feb. 2024).

\bibitem{Vigan2015}
{Vigan}, A., {Gry}, C., {Salter}, G., {Mesa}, D., {Homeier}, D., {Moutou}, C., and {Allard}, F., ``{High-contrast imaging of Sirius A with VLT/SPHERE: looking for giant planets down to one astronomical unit},'' {\em MNRAS}~{\bf 454},  129--143 (Nov. 2015).

\bibitem{Rajan2017}
{Rajan}, A., {Rameau}, J., {De Rosa}, R.~J., {Marley}, M.~S., {Graham}, J.~R., {Macintosh}, B., {Marois}, C., {Morley}, C., {Patience}, J., {Pueyo}, L., {Saumon}, D., {Ward-Duong}, K., {Ammons}, S.~M., {Arriaga}, P., {Bailey}, V.~P., {Barman}, T., {Bulger}, J., {Burrows}, A.~S., {Chilcote}, J., {Cotten}, T., {Czekala}, I., {Doyon}, R., {Duch{\^e}ne}, G., {Esposito}, T.~M., {Fitzgerald}, M.~P., {Follette}, K.~B., {Fortney}, J.~J., {Goodsell}, S.~J., {Greenbaum}, A.~Z., {Hibon}, P., {Hung}, L.-W., {Ingraham}, P., {Johnson-Groh}, M., {Kalas}, P., {Konopacky}, Q., {Lafreni{\`e}re}, D., {Larkin}, J.~E., {Maire}, J., {Marchis}, F., {Metchev}, S., {Millar-Blanchaer}, M.~A., {Morzinski}, K.~M., {Nielsen}, E.~L., {Oppenheimer}, R., {Palmer}, D., {Patel}, R.~I., {Perrin}, M., {Poyneer}, L., {Rantakyr{\"o}}, F.~T., {Ruffio}, J.-B., {Savransky}, D., {Schneider}, A.~C., {Sivaramakrishnan}, A., {Song}, I., {Soummer}, R., {Thomas}, S., {Vasisht}, G., {Wallace}, J.~K., {Wang}, J.~J., {Wiktorowicz}, S., and {Wolff}, S.,
  ``{Characterizing 51 Eri b from 1 to 5 {\ensuremath{\mu}}m: A Partly Cloudy Exoplanet},'' {\em AJ}~{\bf 154},  10 (July 2017).

\bibitem{Ruffio2017}
{Ruffio}, J.-B., {Macintosh}, B., {Wang}, J.~J., {Pueyo}, L., {Nielsen}, E.~L., {De Rosa}, R.~J., {Czekala}, I., {Marley}, M.~S., {Arriaga}, P., {Bailey}, V.~P., {Barman}, T., {Bulger}, J., {Chilcote}, J., {Cotten}, T., {Doyon}, R., {Duch{\^e}ne}, G., {Fitzgerald}, M.~P., {Follette}, K.~B., {Gerard}, B.~L., {Goodsell}, S.~J., {Graham}, J.~R., {Greenbaum}, A.~Z., {Hibon}, P., {Hung}, L.-W., {Ingraham}, P., {Kalas}, P., {Konopacky}, Q., {Larkin}, J.~E., {Maire}, J., {Marchis}, F., {Marois}, C., {Metchev}, S., {Millar-Blanchaer}, M.~A., {Morzinski}, K.~M., {Oppenheimer}, R., {Palmer}, D., {Patience}, J., {Perrin}, M., {Poyneer}, L., {Rajan}, A., {Rameau}, J., {Rantakyr{\"o}}, F.~T., {Savransky}, D., {Schneider}, A.~C., {Sivaramakrishnan}, A., {Song}, I., {Soummer}, R., {Thomas}, S., {Wallace}, J.~K., {Ward-Duong}, K., {Wiktorowicz}, S., and {Wolff}, S., ``{Improving and Assessing Planet Sensitivity of the GPI Exoplanet Survey with a Forward Model Matched Filter},'' {\em ApJ}~{\bf 842},  14 (June 2017).

\bibitem{Stolker2024}
{Stolker}, T., {Kammerer}, J., {Benisty}, M., {Blakely}, D., {Johnstone}, D., {Sitko}, M.~L., {Berger}, J.~P., {Sanchez-Bermudez}, J., {Garufi}, A., {Lacour}, S., {Cantalloube}, F., and {Chauvin}, G., ``{Searching for low-mass companions at small separations in transition disks with aperture masking interferometry},'' {\em A\&A}~{\bf 682},  A101 (Feb. 2024).

\bibitem{Pourre2024}
{Pourr{\'e}}, N., {Winterhalder}, T.~O., {Le Bouquin}, J.~B., {Lacour}, S., {Bidot}, A., {Nowak}, M., {Maire}, A.~L., {Mouillet}, D., {Babusiaux}, C., {Woillez}, J., {Abuter}, R., {Amorim}, A., {Asensio-Torres}, R., {Balmer}, W.~O., {Benisty}, M., {Berger}, J.~P., {Beust}, H., {Blunt}, S., {Boccaletti}, A., {Bonnefoy}, M., {Bonnet}, H., {Bordoni}, M.~S., {Bourdarot}, G., {Brandner}, W., {Cantalloube}, F., {Caselli}, P., {Charnay}, B., {Chauvin}, G., {Chavez}, A., {Choquet}, E., {Christiaens}, V., {Cl{\'e}net}, Y., {Coud{\'e} du Foresto}, V., {Cridland}, A., {Davies}, R., {Defr{\`e}re}, D., {Dembet}, R., {Dexter}, J., {Drescher}, A., {Duvert}, G., {Eckart}, A., {Eisenhauer}, F., {F{\"o}ster Schreiber}, N.~M., {Garcia}, P., {Garcia Lopez}, R., {Gendron}, E., {Genzel}, R., {Gillessen}, S., {Girard}, J.~H., {Gonte}, F., {Grant}, S., {Haubois}, X., {Hei{\ss}el}, G., {Henning}, T., {Hinkley}, S., {Hippler}, S., {H{\"o}nig}, S.~F., {Houll{\'e}}, M., {Hubert}, Z., {Jocou}, L., {Kammerer}, J., {Kenworthy}, M.,
  {Keppler}, M., {Kervella}, P., {Kreidberg}, L., {Kurtovic}, N.~T., {Lagrange}, A.~M., {Lapeyr{\`e}re}, V., {Lutz}, D., {Mang}, F., {Marleau}, G.~D., {M{\'e}rand}, A., {Millour}, F., {Molli{\`e}re}, P., {Monnier}, J.~D., {Mordasini}, C., {Nasedkin}, E., {Oberti}, S., {Ott}, T., {Otten}, G.~P.~L., {Paladini}, C., {Paumard}, T., {Perraut}, K., {Perrin}, G., {Pfuhl}, O., {Pueyo}, L., {Ribeiro}, D.~C., {Rickman}, E., {Rustamkulov}, Z., {Shangguan}, J., {Shimizu}, T., {Sing}, D., {Soulez}, F., {Stadler}, J., {Stolker}, T., {Straub}, O., {Straubmeier}, C., {Sturm}, E., {Sykes}, C., {Tacconi}, L.~J., {van Dishoeck}, E.~F., {Vigan}, A., {Vincent}, F., {von Fellenberg}, S.~D., {Wang}, J., {Widmann}, F., {Yazici}, S., {the GRAVITY Collaboration}, {Abad}, J.~A., {Aller Carpentier}, E., {Alonso}, J., {Andolfato}, L., {Barriga}, P., {Beuzit}, J.~L., {Bourget}, P., {Brast}, R., {Caniguante}, L., {Cottalorda}, E., {Darr{\'e}}, P., {Delabre}, B., {Delboulb{\'e}}, A., {Delplancke-Str{\"o}bele}, F., {Donaldson}, R., {Dorn},
  R., {Dupuy}, C., {Egner}, S., {Fischer}, G., {Frank}, C., {Fuenteseca}, E., {Gitton}, P., {Guerlet}, T., {Guieu}, S., {Gutierrez}, P., {Haguenauer}, P., {Haimerl}, A., {Heritier}, C.~T., {Huber}, S., {Hubin}, N., {Jolley}, P., {Kirchbauer}, J.~P., {Kolb}, J., {Kosmalski}, J., {Krempl}, P., {Le Louarn}, M., {Lilley}, P., {Lopez}, B., {Magnard}, Y., {Mclay}, S., {Meilland}, A., {Meister}, A., {Moulin}, T., {Pasquini}, L., {Paufique}, J., {Percheron}, I., {Pettazzi}, L., {Phan}, D., {Pirani}, W., {Quentin}, J., {Rakich}, A., {Ridings}, R., {Reyes}, J., {Rochat}, S., {Schmid}, C., {Schuhler}, N., {Shchekaturov}, P., {Seidel}, M., {Soenke}, C., {Stadler}, E., {Stephan}, C., {Su{\'a}rez}, M., {Todorovic}, M., {Valdes}, G., {Verinaud}, C., {Zins}, G., {Z{\'u}{\~n}iga-Fern{\'a}ndez}, S., and {the NAOMI Collaboration}, ``{High contrast at short separation with VLTI/GRAVITY: Bringing Gaia companions to light},'' {\em arXiv e-prints} ,  arXiv:2406.04003 (June 2024).

\bibitem{Carter2023}
{Carter}, A.~L., {Hinkley}, S., {Kammerer}, J., {Skemer}, A., {Biller}, B.~A., {Leisenring}, J.~M., {Millar-Blanchaer}, M.~A., {Petrus}, S., {Stone}, J.~M., {Ward-Duong}, K., {Wang}, J.~J., {Girard}, J.~H., {Hines}, D.~C., {Perrin}, M.~D., {Pueyo}, L., {Balmer}, W.~O., {Bonavita}, M., {Bonnefoy}, M., {Chauvin}, G., {Choquet}, E., {Christiaens}, V., {Danielski}, C., {Kennedy}, G.~M., {Matthews}, E.~C., {Miles}, B.~E., {Patapis}, P., {Ray}, S., {Rickman}, E., {Sallum}, S., {Stapelfeldt}, K.~R., {Whiteford}, N., {Zhou}, Y., {Absil}, O., {Boccaletti}, A., {Booth}, M., {Bowler}, B.~P., {Chen}, C.~H., {Currie}, T., {Fortney}, J.~J., {Grady}, C.~A., {Greebaum}, A.~Z., {Henning}, T., {Hoch}, K. K.~W., {Janson}, M., {Kalas}, P., {Kenworthy}, M.~A., {Kervella}, P., {Kraus}, A.~L., {Lagage}, P.-O., {Liu}, M.~C., {Macintosh}, B., {Marino}, S., {Marley}, M.~S., {Marois}, C., {Matthews}, B.~C., {Mawet}, D., {McElwain}, M.~W., {Metchev}, S., {Meyer}, M.~R., {Molliere}, P., {Moran}, S.~E., {Morley}, C.~V., {Mukherjee}, S.,
  {Pantin}, E., {Quirrenbach}, A., {Rebollido}, I., {Ren}, B.~B., {Schneider}, G., {Vasist}, M., {Worthen}, K., {Wyatt}, M.~C., {Briesemeister}, Z.~W., {Bryan}, M.~L., {Calissendorff}, P., {Cantalloube}, F., {Cugno}, G., {De Furio}, M., {Dupuy}, T.~J., {Factor}, S.~M., {Faherty}, J.~K., {Fitzgerald}, M.~P., {Franson}, K., {Gonzales}, E.~C., {Hood}, C.~E., {Howe}, A.~R., {Kuzuhara}, M., {Lagrange}, A.-M., {Lawson}, K., {Lazzoni}, C., {Lew}, B. W.~P., {Liu}, P., {Llop-Sayson}, J., {Lloyd}, J.~P., {Martinez}, R.~A., {Mazoyer}, J., {Palma-Bifani}, P., {Quanz}, S.~P., {Redai}, J.~A., {Samland}, M., {Schlieder}, J.~E., {Tamura}, M., {Tan}, X., {Uyama}, T., {Vigan}, A., {Vos}, J.~M., {Wagner}, K., {Wolff}, S.~G., {Ygouf}, M., {Zhang}, X., {Zhang}, K., and {Zhang}, Z., ``{The JWST Early Release Science Program for Direct Observations of Exoplanetary Systems I: High-contrast Imaging of the Exoplanet HIP 65426 b from 2 to 16 {\ensuremath{\mu}}m},'' {\em ApJL}~{\bf 951},  L20 (July 2023).

\bibitem{Zhang2021}
{Zhang}, Y., {Snellen}, I. A.~G., {Bohn}, A.~J., {Molli{\`e}re}, P., {Ginski}, C., {Hoeijmakers}, H.~J., {Kenworthy}, M.~A., {Mamajek}, E.~E., {Meshkat}, T., {Reggiani}, M., and {Snik}, F., ``{The $^{13}$CO-rich atmosphere of a young accreting super-Jupiter},'' {\em Nature}~{\bf 595},  370--372 (July 2021).

\bibitem{Ruffio2021}
{Ruffio}, J.-B., {Konopacky}, Q.~M., {Barman}, T., {Macintosh}, B., {Hoch}, K. K.~W., {De Rosa}, R.~J., {Wang}, J.~J., {Czekala}, I., and {Marois}, C., ``{Deep Exploration of the Planets HR 8799 b, c, and d with Moderate-resolution Spectroscopy},'' {\em AJ}~{\bf 162},  290 (Dec. 2021).

\bibitem{Ruffio2023}
{Ruffio}, J.-B., {Perrin}, M.~D., {Hoch}, K. K.~W., {Kammerer}, J., {Konopacky}, Q.~M., {Pueyo}, L., {Madurowicz}, A., {Rickman}, E., {Theissen}, C.~A., {Agrawal}, S., {Greenbaum}, A.~Z., {Miles}, B.~E., {Barman}, T.~S., {Balmer}, W.~O., {Llop-Sayson}, J., {Girard}, J.~H., {Rebollido}, I., {Soummer}, R., {Allen}, N.~H., {Anderson}, J., {Beichman}, C.~A., {Bellini}, A., {Bryden}, G., {Espinoza}, N., {Glidden}, A., {Huang}, J., {Lewis}, N.~K., {Libralato}, M., {Louie}, D.~R., {Sohn}, S.~T., {Seager}, S., {van der Marel}, R.~P., {Wakeford}, H.~R., {Watkins}, L.~L., {Ygouf}, M., and {Mountai}, C.~M., ``{JWST-TST High Contrast: Achieving direct spectroscopy of faint substellar companions next to bright stars with the NIRSpec IFU},'' {\em arXiv e-prints} ,  arXiv:2310.09902 (Oct. 2023).

\bibitem{Vigan2024}
{Vigan}, A., {El Morsy}, M., {Lopez}, M., {Otten}, G.~P.~P.~L., {Garcia}, J., {Costes}, J., {Muslimov}, E., {Viret}, A., {Charles}, Y., {Zins}, G., {Murray}, G., {Costille}, A., {Paufique}, J., {Seemann}, U., {Houll{\'e}}, M., {Anwand-Heerwart}, H., {Phillips}, M., {Abinanti}, A., {Balard}, P., {Baraffe}, I., {Benedetti}, J.~A., {Blanchard}, P., {Blanco}, L., {Beuzit}, J.~L., {Choquet}, E., {Cristofari}, P., {Desidera}, S., {Dohlen}, K., {Dorn}, R., {Ely}, T., {Fuenteseca}, E., {Garcia}, N., {Jaquet}, M., {Jaubert}, F., {Kasper}, M., {Le Merrer}, J., {Maire}, A.~L., {N'Diaye}, M., {Pallanca}, L., {Popovic}, D., {Pourcelot}, R., {Reiners}, A., {Rochat}, S., {Sehim}, C., {Schmutzer}, R., {Smette}, A., {Tchoubaklian}, N., {Tomlinson}, P., and {Valenzuela Soto}, J., ``{First light of VLT/HiRISE: High-resolution spectroscopy of young giant exoplanets},'' {\em A\&A}~{\bf 682},  A16 (Feb. 2024).

\bibitem{Kotani2020}
{Kotani}, T., {Kawahara}, H., {Ishizuka}, M., {Jovanovic}, N., {Vievard}, S., {Lozi}, J., {Sahoo}, A., {Guyon}, O., {Yoneta}, K., and {Tamura}, M., ``{Extremely high-contrast, high spectral resolution spectrometer REACH for the Subaru Telescope},'' in [{\em Society of Photo-Optical Instrumentation Engineers (SPIE) Conference Series}{\nolinebreak\hspace{0.1em}]},  {\em Society of Photo-Optical Instrumentation Engineers (SPIE) Conference Series} {\bf 11448},  1144878 (Dec. 2020).

\end{thebibliography}
\bibliographystyle{spiebib} 

\end{document}